\documentclass[english,pra,superscriptaddress,twocolumn]{revtex4-1}
\usepackage{amsmath}
\usepackage{amssymb}
\usepackage{graphicx}
\usepackage{times}

\begin{document}

\title{Perturbative dissipation dynamics of a weakly driven Jaynes-Cummings
system}

\author{S. M. Yu}

\affiliation{School of Applied Sciences, Beijing Univertiy of Technology, Beijing,
100124, China}

\author{Y. B. Gao}

\affiliation{School of Applied Sciences, Beijing Univertiy of Technology, Beijing,
100124, China}

\author{H. Ian}

\affiliation{Institute of Applied Physics and Materials Engineering, University
of Macau, Macau}

\affiliation{UMacau Research Institute, Zhuhai, Guangdong, China}
\begin{abstract}
We generalize a microscopic master equation method to study the dissipation
dynamics of Jaynes-Cummings two-level system with a weak external
driving. Using perturbative analysis to extend the damping bases theory,
we derive the corrected Rabi oscillation and vaccum Rabi splitting
analytically. The evolution of the decoherence factor of the weakly
driven system reveals that the off-diagonal density matrix elements
are oscillating at a frequency dependent on the driving strength and
the initial population inversion. For highly-inverted systems at the
weak-driving limit, this frequency reduces to twice the value for
the non-driven system, showing the dissipation dynamics unable to
be discovered using more conventional approaches.
\end{abstract}
\maketitle

\section{introduction}

The celebrated Jaynes-Cummings (JC) model~\cite{jaynes63} predicts
the splitting of the energy levels of two-level atoms under their
coupling to a quantum field and provide an analytical basis for the
full-quantum oscillation dynamics of two-level systems. The simple
yet elegant model has paved the way to the much ramified research
subject of cavity quantum electrodynamics (cQED)~\cite{Raimond01}
for the atom-field interaction in optical cavities. Recently, it is
incarnated in the cQED studies of superconducting circuit cavities~\cite{wallraff04,majer07}
where the two-level system is a superconducting qubit comprising two
macroscopic persistent-current states, which has contingent applications
in quantum logic gate implementation~\cite{devoret13} for quantum
computation~\cite{divincenzo00}.

It is well-known that one of the major challenges in realizing a functional
quantum computer is the decoherence inherent to a superconducting
qubit, whose sources are the multiple environmental couplings such
as the electron-phonon coupling and the electron-electron coupling
that results in 1/f noises and phase noises innate to a solid-state
system. Therefore, it is imperative to describe the qubit dynamics
that accounts for the decoherence effects more accurately under the
JC-model.

Earlier studies that accounts the decoherence effects employ the so-called
phenomenological master equation that extends the quantum Liouvill
equation for the density matrix to include terms consisting of products
of the density matrix and operators from a semigroup. It is the simplest
implementation of Lindblad's model\cite{lindblad76} for density matrix
evolution in the sense that the coefficients of these terms responsible
for decoherences are assumed a uniform constant $\gamma$ extracted
phenomenologically from experiments. Later, progresses in the direction
of nonequilibrium thermodynamics~\cite{davies74,spohn78} has accumulated
to the Kubo-Martin-Schwinger condition that permits the depiction
of a state-dependent decay coefficient $\gamma(\omega)$ for different
terms in the master equation, where $\omega$ is the eigenfrequency
associated with the operators in the semigroup for decoherence dynamics.
Consequently, the set of decoherence operators can be expanded under
an exact basis of the environmental Hilbert space pertinent to a particular
form of heat-bath interaction Hamiltonian, giving rise to a precise
microscopic master equation derived exactly from a given environmental
coupling~\cite{BREUER}.

Here, we employ the microscopic master equation to study the quantum
dynamics of a cavity QED system with the cavity mode coupled to a
two-level system under the JC-model. The cavity itself is weakly driven
by a resonant laser field. The decay dynamics without an external
driving previously studied by Scala \emph{et al.} shows that the populations
in the quantized levels are oscillating around an averaged exponential
curve derived from the phenomenological theory. When the cavity is
weakly driven, we amend the exactly diagonalizing eigenvectors with
perturbative terms to derive a set of nine damping bases for the Liouville
superoperator, under which decay sources previously underdiscovered~\cite{gao13,gao16,gao09}
are found. Using these damping bases, we find minute-scale oscillations
of higher frequency on top of the original Rabi oscillations~\cite{eberly80,eberly81,rempe87}
that is envelopped by a decay due to the relaxation of the two-level
system. In addition, the splitting of the dressed levels at vacuum
is contracted compared to the non-driven case. More importantly, we
observe the decoherence factor would follow different oscillation
frequencies under different initial population inversions, which can
be explained by the distinct damping bases dominating under particular
inversion cases.

We first derive the microscopic master equation and the associated
eigenoperators in Sec.~\ref{sec:master-equation}. Using these operators
as bases, we compute the correlation functions of the cavity mode
quadrature to show the effective decay rates and the noise density
spectrum in Sec.~\ref{sec:Rabi_oscillation}. The evolution of the
level populations is shown in Sec.~\ref{sec:decoherence} and conclusions
are given in Sec.~\ref{sec:Conclusion}.

\section{microscopic master equation\label{sec:master-equation}}

We start the discussion with the model Hamiltonian $H=H_{\mathrm{JC}}+H_{\mathrm{d}}$
where ($\hbar=1$)
\begin{equation}
H_{\mathrm{JC}}=\omega_{c}a^{\dagger}a+\frac{\omega_{z}}{2}\sigma_{z}+\Omega\left(a\sigma_{+}+a^{\dagger}\sigma_{-}\right)\label{eq:JC_Ham}
\end{equation}
stands for the JC-model Hamiltonian for one two-level system of eigenfrequency
$\omega_{z}$ and one cavity quantum field of frequency $\omega_{c}$
under the rotating wave approximation for their interaction. The weak
external driving is represented by the Hamiltonian
\begin{equation}
H_{\mathrm{d}}=\xi\left(a+a^{\dagger}\right).\label{eq:driving_Ham}
\end{equation}

Under weak-excitation, we consider only the lowest three eigen-levels
for $H_{\mathrm{JC}}$ at atom-cavity resonance: $\vert E_{0}^{\left(0\right)}\rangle=\vert g,0\rangle$
and $\vert E_{\pm}^{\left(0\right)}\rangle=\left(\vert g,1\rangle\pm\vert e,0\rangle\right)/\sqrt{2}$.
Further considering $H_{\mathrm{d}}$ as a weak perturbation to $H_{\mathrm{JC}}$,
we can find the perturbed eigen-basis vectors 
\begin{align}
\vert E_{0}\rangle & =\vert E_{0}^{\left(0\right)}\rangle-\frac{\xi}{\sqrt{2}(\omega_{z}-\Omega)}\vert E_{-}^{\left(0\right)}\rangle-\frac{\xi}{\sqrt{2}(\omega_{z}+\Omega)}\vert E_{+}^{\left(0\right)}\rangle,\label{eq:perVec_0}\\
\vert E_{\pm}\rangle & =\vert E_{\pm}^{\left(0\right)}\rangle+\frac{\xi}{\sqrt{2}(\omega_{z}\pm\Omega)}\vert E_{0}^{\left(0\right)}\rangle,\label{eq:perVec_pm}
\end{align}
in terms of the bare basis vectors under the first-order perturbation
expansion. The perturbation contributes a quadratic term correction
to the eigenvalues from the driveless JC-model, which gives 
\begin{align}
E_{0} & =-\frac{\omega_{z}}{2}-\frac{\omega_{z}\xi^{2}}{\omega_{z}^{2}-\Omega^{2}},\label{eq:perVal_0}\\
E_{\pm} & =\frac{\omega_{z}}{2}\pm\Omega+\frac{\xi^{2}}{2(\omega_{z}\pm\Omega)}\label{eq:perVal_pm}
\end{align}

Pairing the vectors of Eqs.~(\ref{eq:perVec_0})-(\ref{eq:perVec_pm})
with their conjugates form a set of nine damping bases~\cite{briegel93,scala07}
for the Liouville operator of the master equation~\cite{BREUER}
\begin{equation}
\frac{d\rho}{dt}=\mathcal{L}\rho=-i\left[H,\rho\right]+\left(\mathcal{L}_{-}+\mathcal{L}_{+}\right)\rho\label{eq:master_eqn}
\end{equation}
where we have split the environmental contribution into two parts
with each having the effect
\begin{multline}
\mathcal{L}_{\pm}\rho=\frac{\gamma(E_{\pm}-E_{0})}{2}\times\\
\left[\vert E_{0}\rangle\langle E_{\pm}\vert\rho\vert E_{\pm}\rangle\langle E_{0}\vert-\frac{1}{2}\left\{ \vert E_{\pm}\rangle\langle E_{\pm}\vert,\rho\right\} \right]\label{eq:relax_terms}
\end{multline}
on the density matrix. The damping rate $\gamma(\Omega)$ obeys the
Kubo-Martin-Schwinger (KMS) condition $\gamma(-\Omega)=\exp\left\{ -\Omega/k_{B}T\right\} \gamma(\Omega)$
for the quasi-equilibrium system at temperature $T$. Using $\rho_{00}=\left|E_{0}\right\rangle \left\langle E_{0}\right|$
and $\rho_{\alpha\beta}=\left|E_{\alpha}\right\rangle \left\langle E_{\beta}\right|-\delta_{\alpha\beta}\left|E_{0}\right\rangle \left\langle E_{0}\right|$
for all other $\alpha,\beta\in\{0,+,-\}$ to denote the nine damping
bases, the corresponding eigenvalues $\lambda_{\alpha\beta}$ associated
with the eigen-equation $\mathcal{L}\rho_{\alpha\beta}=\lambda_{\alpha\beta}\rho_{\alpha\beta}$
read
\begin{align}
\lambda_{00} & =0,\quad\lambda_{++}=-\frac{\gamma_{+}}{2},\quad\lambda_{--}=-\frac{\gamma_{-}}{2},\label{eq:damp_eigval}\\
\lambda_{0\pm} & =-\frac{\gamma_{\pm}}{4}+i\left[\omega_{z}\pm\Omega+\frac{3\omega_{z}\mp\Omega}{2(\omega_{z}^{2}-\Omega^{2})}\xi^{2}\right],\;\lambda_{\pm0}=\lambda_{0\pm}^{\ast},\nonumber \\
\lambda_{-+} & =-\frac{\gamma_{+}+\gamma_{-}}{4}+i2\Omega\left(1-\frac{\xi^{2}}{\omega_{z}^{2}-\Omega^{2}}\right),\;\lambda_{+-}=\lambda_{-+}^{\ast}.\nonumber 
\end{align}
where the abbreviation $\gamma_{\pm}=\gamma(E_{\pm}-E_{0})$ has been
adopted. We note that since the damping basis $\rho_{00}$ has a zero
eigenvalue, it is obviously identical to the steady-state density
matrix $\rho_{ss}=\rho_{00}$ for $e^{\mathcal{L}t}\rho_{ss}=\rho_{ss}$.
Thus, for any initial state $\rho(0)=\sum_{\alpha,\beta}M_{\alpha\beta}\rho_{\alpha\beta}$
expanded in the damping bases, the master equation (\ref{eq:master_eqn})
admits the formal solution
\begin{equation}
\rho(t)=\sum_{\alpha,\beta}M_{\alpha\beta}\exp\{\lambda_{\alpha\beta}t\}\rho_{\alpha\beta}.\label{eq:formal_soln}
\end{equation}

\section{Corrected Rabi oscillation and vacuum splitting\label{sec:Rabi_oscillation}}

The damping bases with their associated eigenvalues (\ref{eq:damp_eigval})
to the master equation (\ref{eq:master_eqn}) leads to a decaying
envelop to the Rabi oscillation of the two-level system. The effect
of the weak driving can be examined in the temporal domain where we
assume the two-level system is fully inverted initially, i.e. $\left|\psi(0)\right\rangle =\left|e,0\right\rangle $.
In the dressed state bases, one can expand the state to first-order
in the perturbative series as 
\begin{equation}
\left|\psi(0)\right\rangle =\frac{1}{\sqrt{2}}\left(\left|E_{+}\right\rangle -\left|E_{-}\right\rangle \right)+\frac{\xi\Omega}{\omega_{z}^{2}-\Omega^{2}}\left|E_{0}\right\rangle ,
\end{equation}
making the corresponding density matrix $\rho(0)$ be expandable in
the damping bases as
\begin{align}
\rho(0)= & \rho_{00}+\frac{1}{2}(\rho_{--}+\rho_{++}-\rho_{+-}-\rho_{-+})\nonumber \\
 & -\frac{\xi\Omega}{\sqrt{2}(\omega_{z}^{2}-\Omega^{2})}(\rho_{0-}-\rho_{0+}+\rho_{-0}-\rho_{+0}).
\end{align}
The excited-state population $P_{e}=\text{tr}\left\{ \vert e\rangle\langle e\vert\rho(t)\right\} $
can then be computed from Eq.~(\ref{eq:formal_soln}) as a function
of the driving strength $\xi$,
\begin{multline}
P_{e}(\xi)=\frac{1}{4}\left(e^{-\gamma_{-}t/2}+e^{-\gamma_{+}t/2}\right)\\
+\frac{1}{2}e^{-(\gamma_{-}+\gamma_{+})t/4}\cos\Delta t+\frac{\xi^{2}\Omega^{2}}{\left(\omega_{z}^{2}-\Omega^{2}\right)^{2}}\\
\times\left[e^{-\gamma_{-}t/4}\cos(\omega_{-}t)+e^{-\gamma_{+}t/4}\cos(\omega_{+}t)\right],\label{eq:Pe_driven}
\end{multline}
where the first term indicates the spontaneous decay from the two
dressed levels $\left|E_{-}\right\rangle $ and $\left|E_{+}\right\rangle $
while the second term indicates the Rabi oscillation of population.
The Rabi frequency $\Delta=E_{+}-E_{-}$ is corrected from its original
$2\Omega$ for a bare system by a correction term $\xi^{2}\Omega/(\omega_{z}^{2}-\Omega^{2})$
due to the driving, which makes the oscillation slower. In addition,
the driving leads to a third term with $\omega_{\pm}=E_{\pm}-E_{0}$,
which breaks the oscillation symmetry of the dressed levels.

\begin{figure}
\includegraphics[bb=30bp 190bp 550bp 580bp,clip,width=8cm]{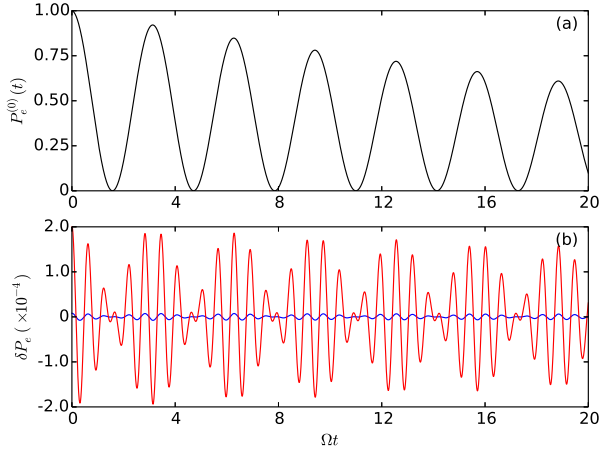}

\caption{(a) Plot of the excited state population $P_{e}^{(0)}$ of a bare
two-level system with no external driving, i.e. $\xi=0$, over dimensionless
time. (b) The change over $P_{e}^{(0)}$ when the system is weakly
driven with driving strength $\xi/\omega_{z}=0.02$ (blue curve) and
$\xi/\omega_{z}=0.1$ (red curve).~\label{fig:Rabi_osc}}
\end{figure}

Figure~(\ref{fig:Rabi_osc}) shows the comparison of the $P_{e}$
evolution over a dimensionless time scale between the non-driven systems
and the driven systems, using the experimental parameters typical
of a superconducting qubit where $\omega_{z}=5$GHz and $\Omega/\omega_{z}=0.2,$
$\gamma_{-}/\omega_{z}=0.002$, and $\gamma_{+}/\omega_{z}=0.006$.
As one expects from Eq.~(\ref{eq:Pe_driven}), the non-driven case
has $P_{e}^{(0)}$ follow an enveloped Rabi oscillation of frequency
$\Delta$ contributed by the second term. The third term proportional
to squared driving strength $\xi^{2}$ has the effect of minute oscillations
of higher frequencies $\omega_{+}$ and $\omega_{-}$ imposed on top
of the Rabi oscillation, which are themselves enveloped by the decay
rates $\gamma_{+}$ and $\gamma_{-}$ respectively. Note that the
summation of the two sinusoidals has the apparent effect that the
minute oscillation has a carrier frequency of $\omega_{+}+\omega_{-}$
and an amplitude oscillation frequency of $\omega_{+}-\omega_{-}=\Delta$.
The first term of Eq.~(\ref{eq:Pe_driven}) contributes to a non-oscillating
offset, which is only visible in the plot over long-term.

To find the spectrum of the driven dressed system, we employ the so-called
input-output theory, whereby the photon statistics of a probed output
field is dependent on the noise input through a correlation Green
function $\left\langle A(\tau)B(0)\right\rangle $ of two operators.
The time-dependent operator is meant to represent $\left\langle A(\tau)\right\rangle =\mathrm{tr}\left\{ e^{\mathcal{L}t}\rho(0)A\right\} $
and the noise density spectrum is the Fourier transform

\begin{equation}
S_{AB}\left(\omega\right)=\frac{1}{\pi}\Re\int_{0}^{\infty}d\tau e^{-i\omega\tau}\langle A\left(\tau\right)B\left(0\right)\rangle.\label{eq:S_AB}
\end{equation}

For the cavity mode considered in the JC-model, we have $A,B\in\{a,a^{\dagger}\}$
so that the resulting correlation functions are essentially the perturbation-corrected
fluctuation-dissipation (FD) relations of the single-mode field under
the framework of microscopic master equation. At the first-order perturbative
expansion given by Eqs.~(\ref{eq:perVec_0})-(\ref{eq:perVec_pm}),
$\left|E_{0}\right\rangle $ is also an eigenvector of $a$, the corrections
contributed by $\left|E_{\pm}\right\rangle $ carrying a negligible
coefficient on the order of $\xi^{2}/(\omega_{c}^{2}-\Omega^{2})$.
Consequently, when we consider the system evolution begin with the
equilibrium state $\rho_{ss}$, it becomes easy to find that
\begin{align}
\langle a^{\dagger}\left(\tau\right)a\left(0\right)\rangle & =\mathrm{tr}\left\{ e^{\mathcal{L}\tau}\vert E_{0}\rangle\langle E_{0}\vert a^{\dagger}a\right\} \nonumber \\
 & =-\frac{\omega_{c}\xi}{\omega_{z}^{2}-\Omega^{2}}\mathrm{tr}\left\{ e^{\lambda_{00}\tau}\left|E_{0}\right\rangle \left\langle E_{0}\right|a\right\} \nonumber \\
 & =\eta^{2},\label{eq:zero_cor}
\end{align}
where $\eta=\xi\omega_{z}/(\omega_{z}^{2}-\Omega^{2})$ is a dimensionless
driving-dependent quantity, which vanishes when the perturbation is
taken to its weak-driving limit. Following the same approach, we can
find 
\begin{equation}
\langle a\left(\tau\right)a\left(0\right)\rangle=\langle a^{\dagger}\left(\tau\right)a^{\dagger}\left(0\right)\rangle=\eta^{2},
\end{equation}
recovering the familiar FD-relations for the cavity vacuum under the
normal approach to dissipation dynamics using phenomenological theories.
The difference stemming from the microscopic approach lies in the
correlation $\langle a\left(\tau\right)a^{\dagger}\left(0\right)\rangle$
because $a^{\dagger}$ is not an eigenvector of $\left|E_{0}\right\rangle $,
rather 
\begin{equation}
a^{\dagger}\left|E_{0}\right\rangle =\frac{1}{\sqrt{2}}\left(\left|E_{+}\right\rangle +\left|E_{-}\right\rangle \right)-\frac{\xi\omega_{z}}{\omega_{z}^{2}-\Omega^{2}}\left|E_{0}\right\rangle .
\end{equation}
Therefore,
\begin{align}
\left\langle a\left(\tau\right)a^{\dagger}\left(0\right)\right\rangle  & =\mathrm{tr}\left\{ e^{\mathcal{L}\tau}\vert E_{0}\rangle\langle E_{0}\vert aa^{\dagger}\right\} \nonumber \\
=\mathrm{tr} & \left\{ \frac{a^{\dagger}}{\sqrt{2}}\left(e^{\lambda_{0+}\tau}\left|E_{0}\right\rangle \left\langle E_{+}\right|+e^{\lambda_{0-}\tau}\left|E_{0}\right\rangle \left\langle E_{-}\right|\right)\right\} \nonumber \\
=\frac{1}{2} & \left[e^{\lambda_{0+}\tau}+e^{\lambda_{0-}\tau}\right]+\eta^{2}\label{eq:nonzero_cor}
\end{align}
where we have omitted the contribution from $\left|E_{0}\right\rangle \left\langle E_{0}\right|$
in the second equality since its coefficient also vanishes in the
weak-driving limit, similar to that of Eq.~(\ref{eq:zero_cor}).

\begin{figure}
\includegraphics[bb=30bp 200bp 594bp 610bp,clip,width=8cm]{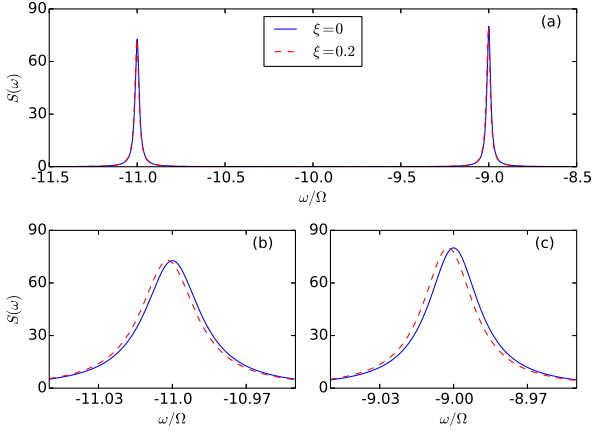}

\caption{(a) Plot of the spectrum $S_{xx}$ for the weak-driving scenario $\xi=0.2\omega_{z}$
(red dashed) and the non-driven scenario $\xi=0$ (blue solid), showing
two peaks of uneven heights and widths at $\omega_{+}$ and $\omega_{-}$.
Magnified views of the left and the right peaks are shown in (b) and
(c), respectively.~\label{fig:Rabi_split}}
\end{figure}

Collecting the four FD-relations above, we have the correlation function
$\langle x\left(\tau\right)x\left(0\right)\rangle$ for the real quadrature
$x=a+a^{\dagger}$ identical to Eq.~(\ref{eq:nonzero_cor}), i.e.
composing of two exponentials. Then it is straightforward to apply
Eq.~(\ref{eq:S_AB}) to find each exponential contribute a Lorentzian
spectral line in the frequency space, i.e.
\begin{multline}
S_{xx}(\omega)=\frac{4\gamma_{-}}{\left(4\omega+4\omega_{-}\right)^{2}+\gamma_{-}^{2}}\\
+\frac{4\gamma_{+}}{\left(4\omega+4\omega_{+}\right)^{2}+\gamma_{+}}+4\eta^{2}\delta(\omega).
\end{multline}
Figure~\ref{fig:Rabi_split} shows a plot of the spectrum where we
have adopted an ohmic dissipation spectrum~\cite{leggett87}
\begin{equation}
\gamma(\omega)=\kappa\omega e^{-\omega/\omega_{\mathrm{C}}}\label{eq:Ohm_bath}
\end{equation}
with cutoff frequency $\omega_{\mathrm{C}}$ and dissipation coefficient
$\kappa$. Then the damping channels along the two levels $\left|E_{+}\right\rangle $
and $\left|E_{-}\right\rangle $ become evidently unequal. With $\omega_{+}>\omega_{-}$
and $\omega_{\mathrm{C}}$ set to 1 GHz, we have $\gamma_{+}<\gamma_{-}$,
making the right Lorentzian at $\omega_{-}$ in the plot exhibit a
higher peak and a wider half-width than the left one at $\omega_{+}$.

The driving results in red-shifts of the two Lorentzians towards to
the left-end of the plot with unequal shift amplitude. As a whole,
the driving as a perturbation has the effect of reducing the vacuum
Rabi split, following the formula
\begin{equation}
\vert\omega_{+}-\omega_{-}\vert=2\Omega\left[1-\frac{\xi^{2}}{2\left(\omega_{z}^{2}-\Omega^{2}\right)}\right].
\end{equation}

\section{Decoherence in Driven Jaynes-Cummings Model\label{sec:decoherence}}

The microscopic master equation approach to the driven JC model also
gives a more accurate prediction of the decoherence dynamics of a
qubit or two-level system in general. Consider an arbitrary initial
state of the qubit and a vacuum state for the resonator, i.e. $\vert\psi\left(0\right)\rangle=c_{g}e^{i\phi}\vert g,0\rangle+c_{e}\vert e,0\rangle,$where
we let $c_{g}$ and $c_{e}$ be real and the complex phase be accounted
by the phase factor $e^{i\phi}$. Under the truncated dressed bases,
the initial state becomes
\begin{multline}
\vert\psi\left(0\right)\rangle=\left(c_{g}e^{i\phi}+\xi\frac{\Omega c_{e}}{\omega_{z}^{2}-\Omega^{2}}\right)\vert E_{0}\rangle\\
+\left(\frac{\xi c_{g}e^{i\phi}}{\omega_{z}-\Omega}-c_{e}\right)\frac{\left|E_{-}\right\rangle }{\sqrt{2}}+\left(\frac{\xi c_{g}e^{i\phi}}{\omega_{z}+\Omega}+c_{e}\right)\frac{\left|E_{+}\right\rangle }{\sqrt{2}}.
\end{multline}
The corresponding pure-state density matrix $\rho=\left|\psi(0)\right\rangle \left\langle \psi(0)\right|$
can be decomposed into the damping bases $\rho=\sum_{\alpha,\beta}M_{\alpha\beta}\rho_{\alpha\beta}$
like we did in last section and here the coefficients are
\begin{align}
M_{00} & =c_{e}c_{g},\\
M_{\pm\pm} & =\frac{1}{2}\left(c_{e}^{2}\pm\frac{2\xi\cos\phi}{\omega_{z}\pm\Omega}c_{e}c_{g}\right),\label{eq:M_pp}\\
M_{0\pm} & =\frac{1}{\sqrt{2}}\left(\pm e^{i\phi}c_{e}c_{g}+\frac{\xi}{\omega_{z}\pm\Omega}c_{g}^{2}\pm\frac{\Omega\xi}{\omega_{z}^{2}-\Omega^{2}}c_{e}^{2}\right),\\
M_{+-} & =\frac{1}{2}\left(-c_{e}^{2}-2\xi\frac{\omega_{z}\cos\phi-i\Omega\sin\phi}{\omega_{z}^{2}-\Omega^{2}}c_{g}c_{e}\right),\label{eq:M_pm}
\end{align}
$M_{\pm0}=M_{0\pm}^{\ast}$, and $M_{-+}=M_{+-}^{\ast}$. 

Consequently, through the formal solution Eq.~(\ref{eq:formal_soln}),
we can find the off-diagonal element $\rho_{eg}=\mathrm{tr}\{\rho\left(t\right)\vert g\rangle\langle e\vert\}$
of the density matrix to be
\begin{align}
\rho_{eg} & =\frac{1}{\sqrt{2}}\left(M_{+0}e^{\lambda_{+0}t}-M_{-0}e^{\lambda_{-0}t}\right)\nonumber \\
 & +\frac{\Omega\xi}{\omega_{z}^{2}-\Omega^{2}}\left(M_{00}e^{\lambda_{00}t}-M_{++}e^{\lambda_{++}t}-M_{--}e^{\lambda_{--}t}\right)\nonumber \\
 & +\frac{\xi}{2(\omega_{z}-\Omega)}\left(M_{+-}e^{\lambda_{+-}t}-M_{--}e^{\lambda_{--}t}\right)\nonumber \\
 & +\frac{\xi}{2(\omega_{z}+\Omega)}\left(M_{++}e^{\lambda_{++}t}-M_{-+}e^{\lambda_{-+}t}\right)\label{eq:rho_eg}
\end{align}
for calculating the decoherence factor $D(t)=|\rho_{eg}|/c_{e}c_{g}$~\cite{zurek05}.
Note that when the driving $\xi$ vanishes, the terms stemmed from
the perturbation in the equation above vanishes and the decoherence
factor reduces to~\cite{gao16}
\begin{multline}
\left[D^{(0)}(t)\right]^{2}=\frac{1}{4}\left(e^{-\gamma_{+}t/4}-e^{-\gamma_{-}t/4}\right)^{2}\\
+e^{-(\gamma_{+}+\gamma_{-})t/4}\cos^{2}(\Omega t),\label{eq:D_0}
\end{multline}
showing that the decoherence in the absence of driving is insensitive
to the initial population distribution in the qubit.

\begin{figure}
\includegraphics[clip,width=8cm]{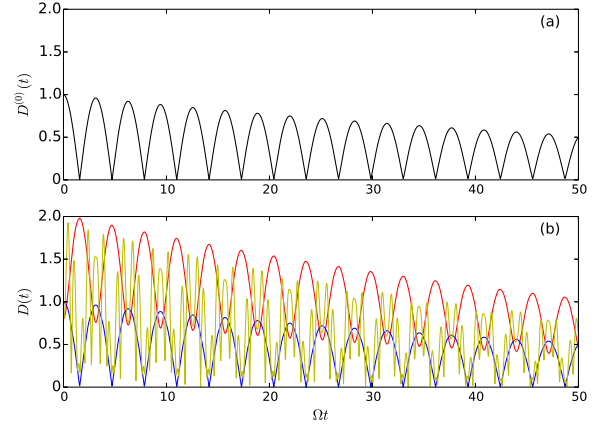}

\caption{(a) Plot of the decoherence factor $D^{(0)}(t)$ as a function of
time in a non-driven two-level system. (b) Plots of the decoherence
factor $D(t)$ under strong coupling $\Omega\approx\omega_{z}$ and
weaking driving $\xi/\omega_{z}=0.1$ when the system is populated
at different proportions: $c_{e}/c_{g}=0.1$ (red curve), $c_{e}/c_{g}=1$
(blue curve), and $c_{e}/c_{g}=100$ (yellow curve).~\label{fig:decoherence}}
\end{figure}

Figure (\ref{fig:decoherence}) shows the comparison of decoherence
factors of the driven cases to the non-driven case, which demonstrates
the inadequacy of the non-driven model to describe the relaxation
dynamics of a driven model where the initial population distribution
plays a crucial role, particularly when the system is strongly coupled
with $\Omega\approx\omega_{z}$. The parameters are typical of superconducting
qubits with $\omega_{z}=5$GHz and $\xi/\omega_{z}=0.1\omega_{z}$.
The decay rates $\gamma_{-}/\omega_{z}=0.05$ and $\gamma_{+}/\omega_{z}=0.055$
are chosen unequal to reflect the uneven damping channels they associated
with for a given environmental reservoir, such as that of Eq.~(\ref{eq:Ohm_bath}).
The influence of population distribution is shown using three different
ratios of $c_{e}/c_{g}$ at $1$ (equal distribution, blue curve),
$0.1$ (ground populated, red curve), and $100$ (excited-state populated,
yellow curve).

We observe that only when the two-level qubit is initially equally
populated would the decoherence factor $D(t)$ resemble that of a
non-driven case. This shows that the driving has minimal effect on
a thermally equilibrated system. On the other hand, when the system
is initially either inverted or condensed to ground, the driving would
set the decoherence into a more complicated oscillation cycle. From
the $c_{e}/c_{g}=0.1$ case, one can see that the grounded population
first experiences a much larger decoherence, about twice, and undergoes
a much quicker decay before it converges with the curve of the non-driven
case. It is also noticeable that the offset, which is mainly due to
the later terms in $M_{\pm0}$, becomes nonzero and the phase of the
decoherence oscillation is inverted from the non-driven case. Since
$c_{e}/c_{g}$ is small, the oscillation is still dominated by the
terms led by the coefficients $M_{\pm0}$ and $M_{00}$, making the
frequency of $D(t)$ similar to that of $D^{(0)}(t)$ given in Eq.~(\ref{eq:D_0}),
i.e. at $\Omega$.

Whereas for the inverted case with $c_{e}/c_{g}$ large, we observe
from Eqs.~(\ref{eq:M_pp})-(\ref{eq:M_pm}) that $\rho_{eg}$ is
dominated by $M_{\pm\pm}$ and $M_{+-}$ terms. Then from their associated
eigenvalues $\lambda_{\pm\pm}$ and $\lambda_{+-}$ of the damping
bases given in Eq.~(\ref{eq:damp_eigval}), the oscillation frequency
becomes close to $2\Omega$. This explains the behavior of the yellow
curve for $c_{e}/c_{g}=100$ in Fig.~\ref{fig:decoherence}.

\begin{figure}
\includegraphics[clip,width=8cm]{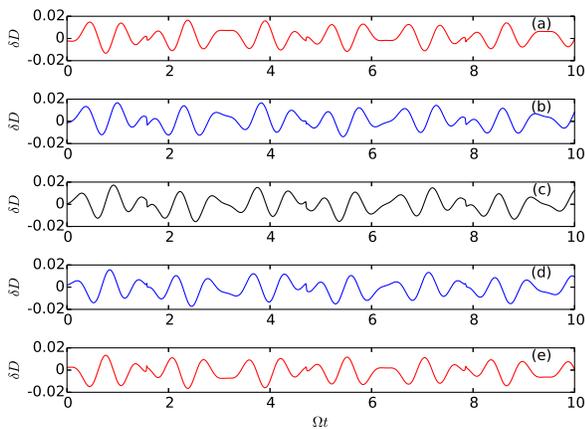}\caption{Plots of the change $\delta D=D(t)-D^{(0)}(t)$ in the decoherence
factor due to different initial phases: (a) $\phi=0$, (b)$\phi=\pi/4$,
(c) $\phi=\pi/2$, (d) $\phi=3\pi/4$, and (e) $\phi=\pi$.~\label{fig:decoh_phase}}
\end{figure}

Moreover, one sees from Eqs.~(\ref{eq:M_pp})-(\ref{eq:M_pm}) that
the complex phase factor $e^{i\phi}$ of the initial state also affects
the decoherence. Fig.~\ref{fig:decoh_phase} shows the change $\delta D$
in the decoherence factor from the non-driven case $D^{(0)}(t)$ when
the phase $\phi$ is set to five values from $0$ to $\pi$. We can
discern clearly a period at $\Omega t=2\pi$ in all subplots and the
inversion of decoherence between a half-period difference, that is
$\delta D(\phi=0)=-\delta D(\phi=\pi)$.

\section{Conclusions and discussions\label{sec:Conclusion}}

We have developed a weakly driven Jaynes-Cummings model under the
microscopic master equation. The effects of the weak drivings are
given as second-order perturbative terms in the eigenvalues associated
with the damping bases of the Liouville superoperator. The exemplified
effects due to the weak driving include corrected Rabi oscillations
and contracted vacuum Rabi splitting. More noticeable is the sensitivity
of the decoherence factor to initial population of the two-level system
given the external driving, which were unaccounted for in previous
non-driven JC-models. Especially when the system is initially inverted,
the decoherence factor would oscillating at twice the Rabi frequency,
showing the decoherence dynamics is dominated by a set of damping
bases different from the usual set for its non-driven counterpart.
This finding demonstrates that the microscopic master equation approach
is not mere corrections to the simpler Langevin equation approach
and Maxwell-Bloch equation approach for analyzing the relaxation process
of a quantum system. On the contrary, it provides a better ground
for analysis that leads to previous undiscovered phenomena.
\begin{acknowledgments}
H. I. acknowledges the support of FDCT Macau under grant 013/2013/A1,
University of Macau under grant MRG022/IH/2013/FST and MYRG2014-00052-FST,
and the National Natural Science Foundation of China under Grant No.~11404415.
\end{acknowledgments}


\begin{thebibliography}{10}
\bibitem{jaynes63} E. T. Jaynes and F. W. Cummings, Proc. IEEE \textbf{51},
89 (1963).

\bibitem{Raimond01} J. M. Raimond, M. Brune, and S. Haroche, Rev.
Mod. Phys. \textbf{73}, 565 (2001).

\bibitem{wallraff04}A. Wallraff, D. I. Schuster, A. Blais, L. Frunzio,
R.-S. Huang, J. Majer, S. Kumar, S. M. Girvin, and R. J. Schoelkopf,
Nature \textbf{431}, 162 (2004). 

\bibitem{majer07}J. Majer, J. M. Chow, J. M. Gambetta, J. Koch, B.
R. Johnson, J. A. Schreier, L. Frunzio, D. I. Schuster, A. A. Houck,
A. Wallraff, A. Blais, M. H. Devoret, S. M. Girvin, and R. J. Schoelkopf,
Nature \textbf{449}, 443 (2007).

\bibitem{devoret13}M. H. Devoret and R. J. Schoelkopf, Science \textbf{339},
1169 (2013).

\bibitem{divincenzo00} D. DiVincenzo, Fortschr. Phys. \textbf{48},
771 (2000) .

\bibitem{lindblad76}G. Lindblad, Commun.Math. Phys. \textbf{48},
119 (1976).

\bibitem{davies74}E. B. Davies, Commun.Math. Phys. \textbf{39}, 91
(1974). 

\bibitem{spohn78}H. Spohn and J. L. Lebowitz, Adv. Chem. Phys. \textbf{38},
109 (1978). 

\bibitem{BREUER} Heinz-Peter Breuer and Francesco Petruccione, \textit{The
Theory of Open Quantum Systems, }(Oxford Press, Oxford 2002).

\bibitem{gao13} C. Chen and Y. B. Gao, Commun. Theor. Phys. \textbf{60},
531 (2013).

\bibitem{gao16} X. Xiao, M. Y. Zhao, S. M. Yu and Y. B. Gao, Commun.
Theor. Phys. \textbf{65}, 273 (2016).

\bibitem{gao09}Y. B. Gao, S. Yang, Yu-xi Liu, C. P. Sun, and Franco
Nori, arxiv: 0902.2512.

\bibitem{eberly80} J. H. Eberly, N. B. Narozny, and J. J. Sanchez-
Mondragon, Phys. Rev. Lett.\textbf{ 44}, 1323 (1980).

\bibitem{eberly81} N. B. Narozny, J. J. Sanchez-Mondragon, and J.
H. Eberly, Phys. Rev. A. \textbf{23}, 236 (1981).

\bibitem{rempe87} G. Rempe, H. Walther, and N. Klein, Phys. Rev.
Lett. \textbf{58}, 353 (1987).

\bibitem{briegel93} H.-J. Briegel and B.-G. Englert, Phys. Rev. A
\textbf{47}, 3311 (1993).

\bibitem{scala07} M. Scala, B. Militello, A. Messina, J. Piilo, S.
Maniscalco, Phys. Rev. A \textbf{75}, 013811 (2007).

\bibitem{leggett87} A. J. Leggett, S. Chakravarty, A. T. Dorsey,
Matthew P. A. Fisher, Anupam Garg, and W. Zwerger, Rev. Mod. Phys.
\textbf{59}, 1 (1987).

\bibitem{zurek05}F. M. Cucchietti, J. P. Paz, and W. H. Zurek, Phys.
Rev. A \textbf{72}, 052113 (2005). 
\end{thebibliography}
\end{document}